\documentclass[a4paper,10pt]{article}
\usepackage[utf8]{inputenc}
\usepackage{mathtools,amsmath,amssymb}
\usepackage{graphicx}
\usepackage{lmodern}
\usepackage[T1]{fontenc}
\usepackage{microtype}
\usepackage[numbers,sort]{natbib}
\usepackage{titlesec}
\usepackage{xspace}
\usepackage[left=1.8cm,right=1.8cm,bottom=2.3cm,top=2cm]{geometry}
\usepackage[font=it, labelfont=normalfont,bf]{caption}
\usepackage{multicol}
\usepackage{float}
\renewcommand{\footnoterule}{%
  \kern 10pt
}

\renewcommand{\thesection}{\Roman{section}}

\titleformat{\section}
{\centering\bf}{\thesection}{1em}{\noindent}

  \titlespacing{\section}{0pt}{2\baselineskip}{1\baselineskip}

\newcommand{\dd}{\mathrm{d}}

\newcommand{\ola}[1]{{\bar{#1}}}

\newcommand{\NN}{n}
\newcommand{\KK}{k}
\newcommand{\amp}{\mathcal{A}}
\newcommand{\cut}{\mathrm{cut}}
\newcommand{\NNcut}{{\NN_{\cut}}}
\newcommand{\planar}{\mathrm{p}}
\newcommand{\NNp}{{\NN_{\planar}}}
\newcommand{\KKp}{{\KK_{\planar}}}
\newcommand{\ampp}{\amp_{\planar}}
\newcommand{\nonplanar}{\mathrm{np}}
\newcommand{\NNnp}{{\NN_{\nonplanar}}}
\newcommand{\KKnp}{{\KK_{\nonplanar}}}
\newcommand{\ampnp}{\amp_{\nonplanar}}
\newcommand{\cylinder}{\mathrm{cyl}}

\newcommand{\KKc}{{\KK_{\cylinder}}}
\newcommand{\ampc}{\amp_{\cylinder}}

\newcommand{\ccc}{\mathbb{C}}
\newcommand{\lax}{\mathcal{L}}
\newcommand{\mono}{\mathcal{M}}
\newcommand{\trans}{\mathcal{T}}
\newcommand{\ww}{\mathcal{W}}
\newcommand{\dw}{\partial}
\newcommand{\oo}{\mathcal{O}}
\newcommand{\eqndot}{\;{}.}
\newcommand{\eqncom}{\;{},}
\DeclareMathOperator{\str}{str}

\begin{document}

    \texttt{\footnotesize
    \strut\hfill DCPT-16/05 \\
    \strut\hfill HU-EP-16/08 \\
    \strut\hfill HU-MATH-2016-04 \\
    }

\vspace{1em}

\begin{center}
 \textbf{\Large Yangian-type symmetries of non-planar leading singularities}
\\\vspace{1em}
{\normalsize Rouven Frassek$^{\,a}$ \, \& \, David Meidinger$^{\,b}$}
\end{center}
{\small
$^{\,a}\,$\textit{Department of Mathematical Sciences, Durham University,
South Road, Durham DH1 3LE, United Kingdom}\\
\hspace*{8pt}\texttt{rouven.frassek@durham.ac.uk}\\
$^{\,b}\,$\textit{Institut für Mathematik und Institut für Physik,
Humboldt-Universität zu Berlin,
Zum Großen Windkanal 6,
12489 Berlin}\\
\hspace*{8pt}\texttt{david.meidinger@physik.hu-berlin.de}
 }
 \vspace{1em}
\begin{center}
    \begin{minipage}{\textwidth}
    \small
 \noindent  
We take up the study of integrable structures behind non-planar contributions
to scattering amplitudes in $\mathcal{N}\! = \! 4$ super Yang-Mills theory. 
Focusing on leading singularities, we derive the action of the Yangian generators on color-ordered subsets of the
external states. Each subset corresponds to a single boundary of the non-planar on-shell diagram.
While Yangian invariance is broken, we find that higher-level Yangian generators still annihilate the non-planar on-shell diagram. 
For a given diagram, the number of these generators is governed by the degree of non-planarity.
Furthermore, we present additional identities involving 
integrable transfer matrices. 
In particular, for diagrams on a cylinder we obtain a conservation rule similar to the Yangian invariance condition of planar on-shell diagrams.
To exemplify our results, we consider a five-point MHV on-shell function on a cylinder.
\end{minipage}
\end{center}
\vspace{1em}

\begin{multicols}{2}

\section{Introduction}\noindent
Over the last decades there was tremendous progress
in the understanding of scattering amplitudes of quantum field theories,
in particular for $\mathcal{N}\! = \! 4$ super Yang-Mills.
For this theory it was found that in addition to superconformal invariance,
tree level amplitudes, as well as planar loop-level integrands exhibit dual 
superconformal symmetry~\cite{Drummond:2008vq}.
Together they combine into a Yangian~\cite{Drummond:2009fd}, a symmetry  that also underlies the integrable structure of the spectral problem~\cite{Beisert:2010jr,Beisert:2010jq}.

More recently, the authors of~\cite{Bern:2014kca,Arkani-Hamed:2014via,Bern:2015ple} observed that for the examples they studied,
non-planar contributions to the integrand exhibit properties, which in the
planar case are consequences of dual superconformal symmetry.
This is a strong hint that planar integrability implies constraints on
the subleading contributions in the $1/N$ expansion.
Here we make a preliminary step towards identifying these symmetries,
focusing on leading singularities instead of the full integrand.

Leading singularities are quantities that are obtained from the loop integrand
by localizing all integration variables by Cutkovsky cuts. 
They capture the IR structure of the amplitude
and play an important role
in the method of generalized unitarity~\cite{Bern:1994zx,Bern:1994cg,Britto:2004nc}.
As discussed in~\cite{ArkaniHamed:2012nw}, all leading singularities can be expressed as on-shell diagrams. 
On-shell diagrams are graphs with black and white trivalent vertices, representing 3-point
$\text{MHV}$ and $\overline{\text{MHV}}$ amplitudes.
Their internal edges correspond to on-shell phase space integrations.
Due to BCFW recursion relations at tree-~\cite{Britto:2004ap,Britto:2005fq} and loop-level~\cite{CaronHuot:2010zt,ArkaniHamed:2010kv}, planar on-shell 
diagrams also encode all tree level scattering amplitudes as well as the integrands of planar
loop-level amplitudes.

In the following we use twistor variables $\ww_i^a$, the half-Fourier transform of spinor-helicity variables~\cite{Witten:2003nn}. However, our results are independent of this choice.
When working with the twistor variables above,
two external legs of an on-shell diagram are glued together
by identifying the legs employing a projective delta function
\begin{equation}\label{eq:line}
    \Delta_{ij} 
    \, = \int\frac{\dd\alpha}{\alpha}\delta^{4|4}(\ww_i+\alpha\ww_j)
    \eqncom
\end{equation}
and subsequently integrating over the internal states 
\begin{equation}\label{eq:gluing}
    \int \dd^{3|4}\ww
    \, =      
    \int\frac{\dd^{4|4}\ww}{\text{Vol}[GL(1)]}
    \eqndot
\end{equation}
Every on-shell diagram encodes an 
expression in terms of a Graßmannian contour
integral of the form~\cite{ArkaniHamed:2009dn,ArkaniHamed:2012nw,Arkani-Hamed:2014bca,Franco:2015rma}
\begin{equation}\label{eq:grassmannian}
    \amp \, = 
    \int \frac{\dd^{\KK\times \NN}{\rm C}}{\text{Vol}[GL(\KK)]} \;\, \Omega \;\,
    \delta^{4\KK | 4\KK}({\rm C}\cdot\ww)
    \eqncom
\end{equation}
where $\Omega$ is a rational function of the $\KK\times\KK$ minors of the $\KK\times\NN$ matrix ${\rm C}$. Here $\KK$ is the MHV degree and $\NN$ the number of particles. 
While the integrand $\Omega$ is fixed for planar diagrams by Yangian invariance and simply given by the inverse of the product of all consecutive minors~\cite{Drummond:2010uq,Drummond:2010qh}, a general expression for non-planar diagrams is unknown and has to be calculated case-by-case~\cite{Gekhtman2009,Franco:2013nwa,Franco:2015rma}.

In this article we derive symmetries of non-planar leading singularities (on-shell diagrams) that are inherited from the Yangian invariance of their planar counterparts. 
Noting that every non-planar on-shell diagram can be cut
open until it is planar allows us to deduce the action of the Yangian generators on a given boundary
with a fixed ordering of external states.
We find that that non-planar leading singularities are
invariant under a subset of the Yangian generators~\eqref{eq:resultyangian}.
The first levels are broken depending on the degree of non-planarity, which is parametrized by the number of cut internal edges.
Additionally, we derive similar identities involving integrable transfer matrices which 
are related to the Yangian \eqref{eq:resulttransfer}.
For the special case of diagrams on a cylinder, we present an exact conservation law in~\eqref{eq:cylinderidentity}. Finally, we consider a five-point MHV on-shell function on a cylinder to exemplify our results.

\section{Monodromy matrix identities}\label{sec:2}
\begin{figure}[H]
    \vspace*{2\baselineskip}
    \centering
    \includegraphics[width=0.9\linewidth]{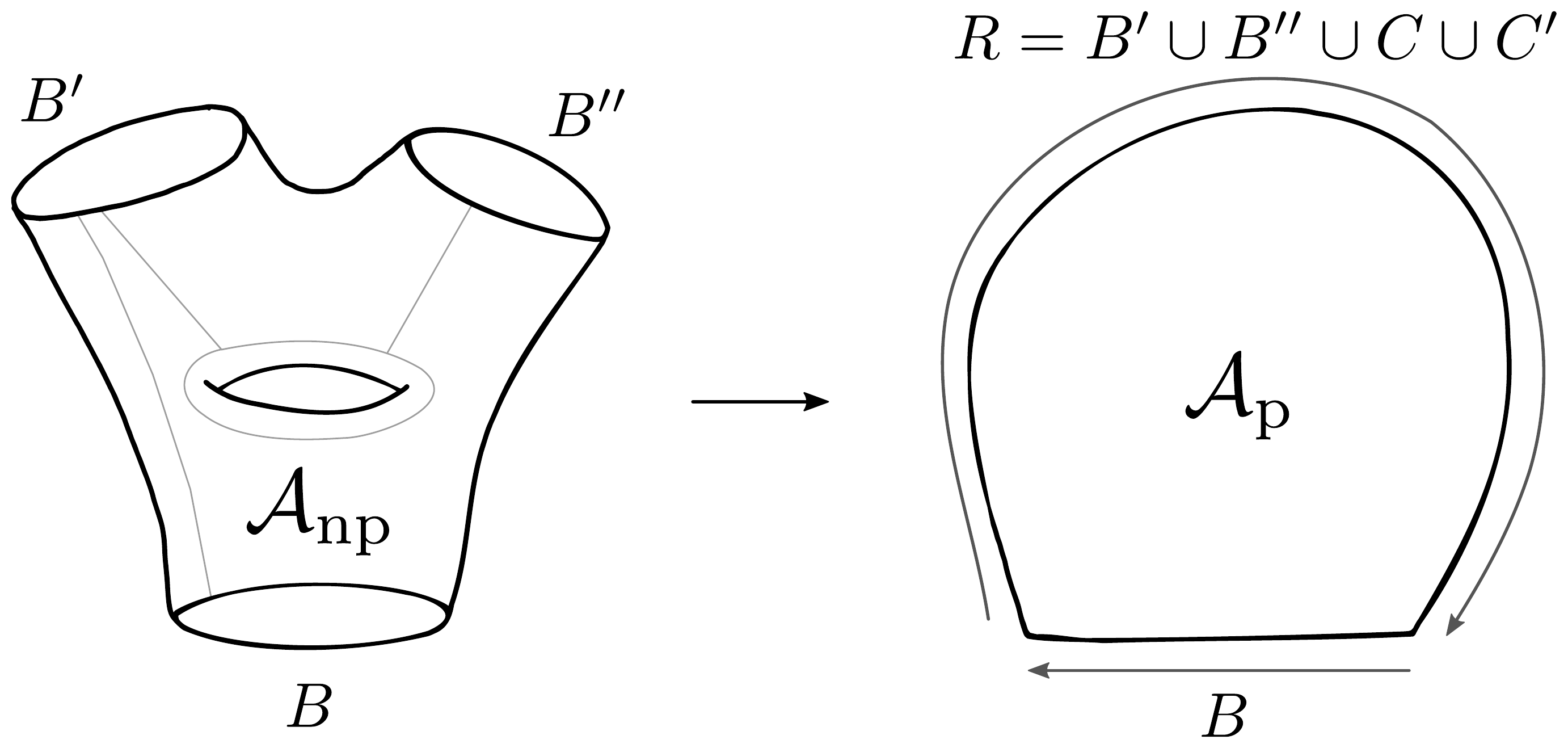}
    \caption{
        Cutting a non-planar into a planar on-shell diagram. 
        We only show the surface on which the diagram is embedded.
        The gray lines indicate a possible cuts.
        For the planar diagram we combine the states on boundaries $B'$ and $B''$
        together with the cut lines $C$ and $C'$ into the ordered set $R$.
        The arrows indicate the ordering of labels.
        \vspace*{-3pt}
    }
    \label{fig:cutting1}
\end{figure}

\noindent
Let us consider an arbitrary non-planar on-shell diagram $\ampnp$, with $\NNnp$ external particles
and MHV degree $\KKnp$.
By cutting internal edges this diagram can be cast into a planar diagram $\ampp$. As a consequence, $\ampnp$ can be written in terms of this diagram $\ampp$ with external states identified
via \eqref{eq:line} and \eqref{eq:gluing},
\begin{equation}\label{eq:nplan}
    \ampnp = 
    \int_{C,C'} 
    \,
    \Delta_{CC'} 
    \;
    \ampp
    \eqndot
\end{equation}
Here $\int_{C,C'}$ is a shorthand notation for the projective integrations \eqref{eq:gluing} over all the cut internal states of $\ampnp$ with 
$\Delta_{CC'}=\Delta_{C_1 C'_1} \cdots \Delta_{C_{\NNcut}C'_{\NNcut}}$ as defined in \eqref{eq:line}.
Schematically, the decomposition \eqref{eq:nplan} is depicted in Figure~\ref{fig:cutting1}.
The planar diagram $\ampp$ has $\NNp=\NNnp+2\NNcut$ external legs and MHV degree $\KKp=\KKnp+\NNcut$. In general the decomposition in \eqref{eq:nplan} is not unique and the parameters $\NNp$ and $\KKp$ depend on this choice. 
In the following, we distinguish among the particles on a single boundary $B$ of 
the non-planar diagram $\ampnp$ and the remaining external particles of $\ampp$ which we label by $R$. They include the particles on other boundaries of $\ampnp$ as well as the $2 \NNcut$ particles that become external when cutting the diagram open, see Figure~\ref{fig:cutting1}.
Initially, Yangian invariance of the planar on-shell diagram $\ampp$ was shown using Drinfeld's first realization~\cite{Drummond:2009fd}. 
In this realization the Yangian is defined via the first and second level generators and their commutators, imposing certain Serre relations.%
\noindent
In the following, we will work in the 
RTT-realization of the Yangian which is intimately related to the construction of spin chain monodromy matrices 
in the framework of the quantum inverse scattering method. In this realization all Yangian generators are combined into such a monodromy matrix, which depends on a spectral parameter. The individual generators of the Yangian symmetry can be recovered by expanding in this parameter.
Yangian invariance of the planar on-shell diagram $\ampp$ can then be compactly expressed as a set of $8\times 8$ eigenvalue equations
\begin{equation}\label{eq:planarinv}
    \mono^{ab}_{RB}(u)\ampp
    = 
    \lambda(u) \, \delta^{ab} \, \ampp
    \eqncom
\end{equation}
see~\cite{Frassek:2013xza,Chicherin:2013ora}. Here the indices take the values $a,b=1,\ldots,8$, the complex variable $u$ denotes the spectral parameter and
the eigenvalue in the conventions fixed below is $\lambda(u)=(u-1)^{\KKp}u^{\NNp-\KKp}$. The monodromy matrix $\mono_{RB}$ can be written as the product of two $8\times 8$ monodromy matrices acting on $B$ and $R$
\begin{equation}\label{eq:monosplit}
    \mono_{RB}(u)= \mono_{R}(u) \, \mono_{B}(u)
    \eqndot
\end{equation}
Each of the monodromies yields a realization of the Yangian separately. They are defined in terms of the Lax operators $\lax_i(u)=u+(-1)^b e^{ab}\, \ww_i^b \dw_i^a$ as
\begin{equation}\label{eq:monolax}
    \mono_{B}(u)
    =
    \lax_{B_1}(u)
    \cdots 
    \lax_{B_{\NN_B}}(u) 
    \eqncom
\end{equation}
with  $\mono_{R}(u)$ correspondingly.
The elementary matrices $e^{ab}$ form the fundamental representation of $\mathfrak{gl}(4|4)$, while the operators $\ww_i^b \dw_i^a$ generate the action of the superconformal group on particle $i$. Here we introduced the notation $\dw_i^a=\partial/\partial \ww_i^a$.

We now show that identities similar to the Yangian invariance of the planar on-shell diagram $\ampp$ in
\eqref{eq:planarinv} also hold for the non-planar diagram $\ampnp$.
In order to derive those identities we note that the product of two Lax operators $\lax_i$ for a certain choice of the spectral parameters is proportional to the identity
\begin{equation}\label{eq:unitarity}
    \lax_i(u)\lax_i(1-u-\ccc_i)
    = 
    u(1-u-\ccc_i )
    \eqndot
\end{equation}
In the context of integrable models this property is also known as unitarity. The central charges $\ccc_i=\ww_i^a\partial_i^a$ vanish 
when acting on an on-shell diagram.
Using the inversion relation in \eqref{eq:unitarity},
the planar Yangian invariance condition \eqref{eq:planarinv}  can be rewritten as
\begin{equation}\label{eq:actiononplanar}
    \mono_{B}(u) \ampp
    = 
    \frac{(-1)^{\KKp} u^{\NN_B-\KKp} }{(1-u)^{\NN_R-\KKp}} \mono_{\ola{R}}(1-u) \ampp
    \eqndot
\end{equation}
Here the Lax operators in $\mono_{\ola{R}}(u)=\lax_{R_{\NN_R}}(u) \cdots \lax_{R_{1}}(u)$
are multiplied in the opposite order compared to \eqref{eq:monolax}.
The monodromy matrix on the left-hand side does not depend on the cut lines. Thus from the definition of the non-planar on-shell diagram $\ampnp$ in \eqref{eq:nplan} we immediately conclude that
\begin{equation}\label{eq:actiononnonplanar}
    \mono_{B}(u) \ampnp
    = 
    \frac{(-1)^{\KKp}u^{\NN_B-\KKp}}{(1-u)^{\NN_R-\KKp}} 
    \int_{C,C'} \Delta_{CC'} 
    \mono_{\ola{R}}(1-u) \ampp
    \eqndot
\end{equation}
To obtain the action of the Yangian generators on the boundary $B$ we expand the monodromy in terms of the spectral parameter $u$. This yields the Yangian generators
\begin{equation}\label{eq:expansion}
    \mono^{ab}_{B}(u)=u^{\NN_B}\delta^{ab}+u^{\NN_B-1}\big(\mono^{[1]}_{B}\big)^{ab}+\cdots+\big(\mono^{[\NN_B]}_{B}\big)^{ab}
    \eqncom
\end{equation} 
see e.g.~\cite{Molev:1994rs}. The form of the Yangian generators $(\mono_{B}{}^{[i]})^{ab}$ can be obtained from the monodromy \eqref{eq:monolax} after inserting the explicit form of the Lax operators. The analogous expansion holds for the monodromy $\mono_{\ola{R}}(u)$. Thus, when expanding \eqref{eq:actiononnonplanar} for $u\ll1$ we obtain the action of the Yangian generators $(\mono_B{}^{[i]})^{ab}$ on the boundary $B$.
We find that the action of the first $\KKp$ Yangian generators is rather complicated 
\begin{equation}\label{eq:resultyangian2}
    \big(\mono^{[i]}_{B}\big)^{ab} 
    \ampnp=\sum_{j=0}^{\NN_R}\frac{(j-\KKp)_{\KKp-i}}{(\KKp-i)!}  \int_{C,C'}\!\!\!\Delta_{CC'}  \big(\mono^{[j]}_{\ola{R}}\big)^{ab}\ampp
\end{equation}
for $i=0,\dots,\KKp$.
Here $(\mono_{\ola{R}}{}^{[j]})^{ab}$ denote the Yangian generators of the monodromy $\mono_{\ola{R}}(u)$ involving
superconformal generators which are inserted into the diagram and $(a)_n$ is the Pochhammer symbol.
However, we find that the remaining higher levels of the Yangian generators that act on the boundary $B$ annihilate the non-planar on-shell diagram $\ampnp$, and generate unbroken symmetries,
\begin{equation}\label{eq:resultyangian}
    \big(\mono^{[i]}_{B}\big)^{ab} 
    \,\, \ampnp \, = \,  0
    \, , \qquad
    i=\KKp+1,\dots,\NN_B
    \eqndot
\end{equation}
Note that the amount of unbroken symmetries in \eqref{eq:resultyangian} depends in an interesting
way on the degree of non-planarity and is given by $\NN_B - \KKp$. The number of external states $\NN_B$ fixes the number of levels of the  Yangian generators
and $\KKp=\KKnp+\NNcut$ can be regarded as a measure of non-planarity,
as each additional boundary or handle requires further internal lines to be cut. 
If $\KKp\leq\NN_B$, we don't find unbroken symmetries.
Here we did not specify any particular embedding of the diagram, nor 
any specific way of cutting it into a planar one. 
The preceding discussion shows that the actual symmetries are determined by the minimal way to cut the diagram,
and that one should consider all possible embeddings of the diagram to identify
as many symmetries as possible.

\section{Transfer matrix identities}\label{sec:3}\noindent
In the spirit of the quantum inverse scattering method~\cite{Faddeev:1996iy}, 
we define the transfer matrix as the supertrace over the auxiliary fundamental space of the monodromy matrix
\begin{equation}\label{eq:deftrans}
    \trans_B(u)=\str\mono_B(u)
    \eqncom
\end{equation}
see also~\cite{Frassek:2015rka}, where this transfer matrix appeared in the context of on-shell diagrams of form factors.
It generates a set of mutually commuting operators $\trans_B^{[i]}=\str \mono_B^{[i]}$ with $i=1,\ldots,\NN_B$, cf.~\eqref{eq:expansion}.
In this section we derive further identities of non-planar on-shell diagrams which involve the operators $\trans_B^{[i]}$. 
In particular, for the special case of diagrams 
on a cylinder they yield exact conservation laws.
\begin{figure}[H]
    \vspace*{3pt}
    \centering
    \includegraphics[width=0.9\linewidth]{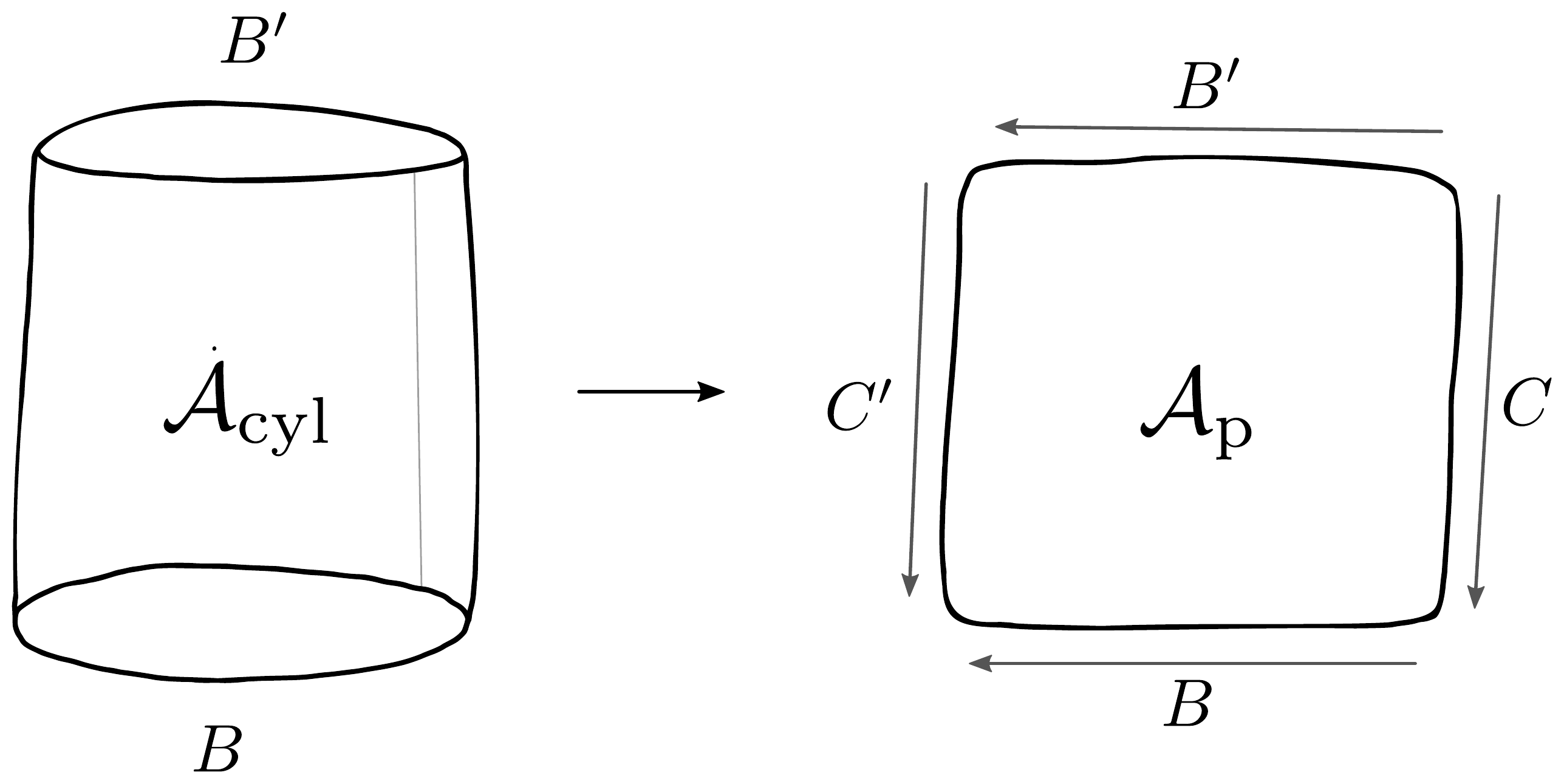}
    \vspace*{2pt}
    \caption{
        Cutting an on-shell diagram on a cylinder into a planar on-shell diagram.
        We only show the surface on which the diagram is embedded.
        The gray line indicates a possible cut.
        The external states $C$ and $C'$ arise from cutting internal lines of $\ampc$.
        Arrows indicate the ordering of labels.
    \vspace*{0pt}
    }
    \label{fig:cutting2}
\end{figure}\noindent

First, we specialize~\eqref{eq:actiononnonplanar} to the case of a diagram with two boundaries. Then the supertrace yields
\begin{multline}\label{eq:actiononlanarcylinder}
    \!\!\!\!\!\trans_{B}(u) \ampc
    =\\[6pt] \,
    \cramped{\frac{(-1)^{\KKp}u^{\NN_B-\KKp}}{(1-u)^{\NN_{B'}+2\NNcut-\KKp}} 
    \int_{C,C'}\!\!\!\!
    \Delta_{CC'} 
    \trans_{\ola{C} B' C'}(1-u) \ampp}
    \eqncom
\end{multline}
where $\ampc$ denotes an on-shell diagram on a cylinder, cf.~Figure~\ref{fig:cutting2}.
At first sight we have not gained anything in comparison to~\eqref{eq:actiononnonplanar}. However, when acting with the transfer matrix on $\ampc$ we can evaluate the integral over the internal lines on the right-hand side.
We first integrate by parts using
\begin{multline}
    \textstyle{\int}
    \dd^{3|4}\ww \, g(\ww) \lax(u) f(\ww)
    \\=
    -
    \textstyle{\int}
    \dd^{3|4}\ww \left[ \lax(1-u) g(\ww) \right] f(\ww)
    \eqncom
\end{multline}
which holds for arbitrary functions $f$ and $g$.
Now the Lax operators $\lax_{C_i}$ and $\lax_{C'_i}$ act on the $\Delta$'s instead of $\ampp$.
The special feature of diagrams on cylinders is that here, the cyclicity of the supertrace
allows to bring the Lax operators $\lax_{C_i}$ and $\lax_{C'_i}$ into a consecutive order, $\trans_{\ola C B' C'} = \trans_{C' \ola C B'}$.
We can now use the identity
\begin{equation}\label{eq:yangianline}
    \lax_{C'_i}(u)\lax_{C_i}(u)\Delta_{C_{i}C'_{i}}
    =u(u-1)\Delta_{C_{i}C'_{i}}
    \eqncom
\end{equation}
which is equivalent to the inversion relation in \eqref{eq:unitarity}.
This removes these Lax operators entirely from the right-hand side,
and the transfer matrix becomes simply $\trans_{B'}(1-u)$ and can be pulled
out of the integral. The integral is then the original diagram on the cylinder,
$\ampc=\int_{CC'}\Delta_{CC'}\ampp$, and we finally find
\begin{equation}\label{eq:cylinderidentity}
    u^\KKc \, \frac{\trans_B(u)}{u^{\NN_B}} \;\, \ampc
    \; = \;
    (u-1)^{\KKc}\, \frac{\trans_{B'}(1-u)}{(1-u)^{\NN_{B'}}} \;\, \ampc
    \eqncom
\end{equation}
This result can be understood as a conservation law or intertwining relation
between the two boundaries.
Comparing with~\eqref{eq:actiononplanar}, we see that that it
plays a similar role as an exact identity for the cylinder as the
Yangian invariance does for planar diagrams.
Note in particular that there is no dependence on the number of cut lines.%

Similar to the case of the monodromy in Section~\ref{sec:2}, we use~\eqref{eq:cylinderidentity} to obtain further identities for
general non-planar on-shell diagrams.
Again we consider an arbitrary on-shell diagram $\ampnp$.
This time we decompose it only up to a diagram on a cylinder such that
\begin{equation}\label{eq:cutcyl}
    \ampnp = 
    \int_{C,C'} 
    \,
    \Delta_{CC'} 
    \;
    \ampc
    \eqndot
\end{equation}
The on-shell diagram on the cylinder $\ampc$ satisfies~\eqref{eq:cylinderidentity}. 
Here we take $B$ to be an actual boundary of $\ampnp$. 
The other boundary $B'$ of $\ampc$ contains the other boundaries of the initial diagram as well as the cut lines $C_i$, $C'_i$.
Integrating this identity over the cut lines as in \eqref{eq:cutcyl} we get
\begin{equation}\label{eq:cylcyl}
    \trans_{B}(u) \ampnp
    = 
    \frac{(-1)^{\KKc}u^{\NN_{B}-\KKc}}{(1-u)^{\NN_{B'}-\KKc}} 
    \int_{C,C'}\!\!\!\!\Delta_{CC'} 
    \trans_{\ola{B'}}(1-u) \ampc
    \eqncom
\end{equation}
By arguments identical to those used in Section~\ref{sec:2}, 
we can expand in the spectral parameter and identify powers 
where the right-hand side of \eqref{eq:cylcyl} vanishes:
\begin{equation}\label{eq:resulttransfer}
    \trans_{B}^{[i]} \; \ampnp \, = \, 0 
    \, , \qquad 
    i=\KKc+1,\dots,\NN_B
    \eqndot
\end{equation}
Note that although~\eqref{eq:resulttransfer} looks
like the supertrace of~\eqref{eq:resultyangian}, 
the crucial difference lies in the number of broken levels:
Here $\KKc$ refers to the MHV degree after cutting to a cylinder,
which is smaller than $\KKp$ the MHV degree after continuing to cut the diagram to a planar one.
Thus~\eqref{eq:resulttransfer} provides additional identities not obtained from the supertrace of~\eqref{eq:resultyangian}.

\section{Example: Five-point MHV on a cylinder}\label{sec:example}
\begin{figure}[H]
    \vspace*{2pt}
    \centering
    \includegraphics[width=0.7\linewidth]{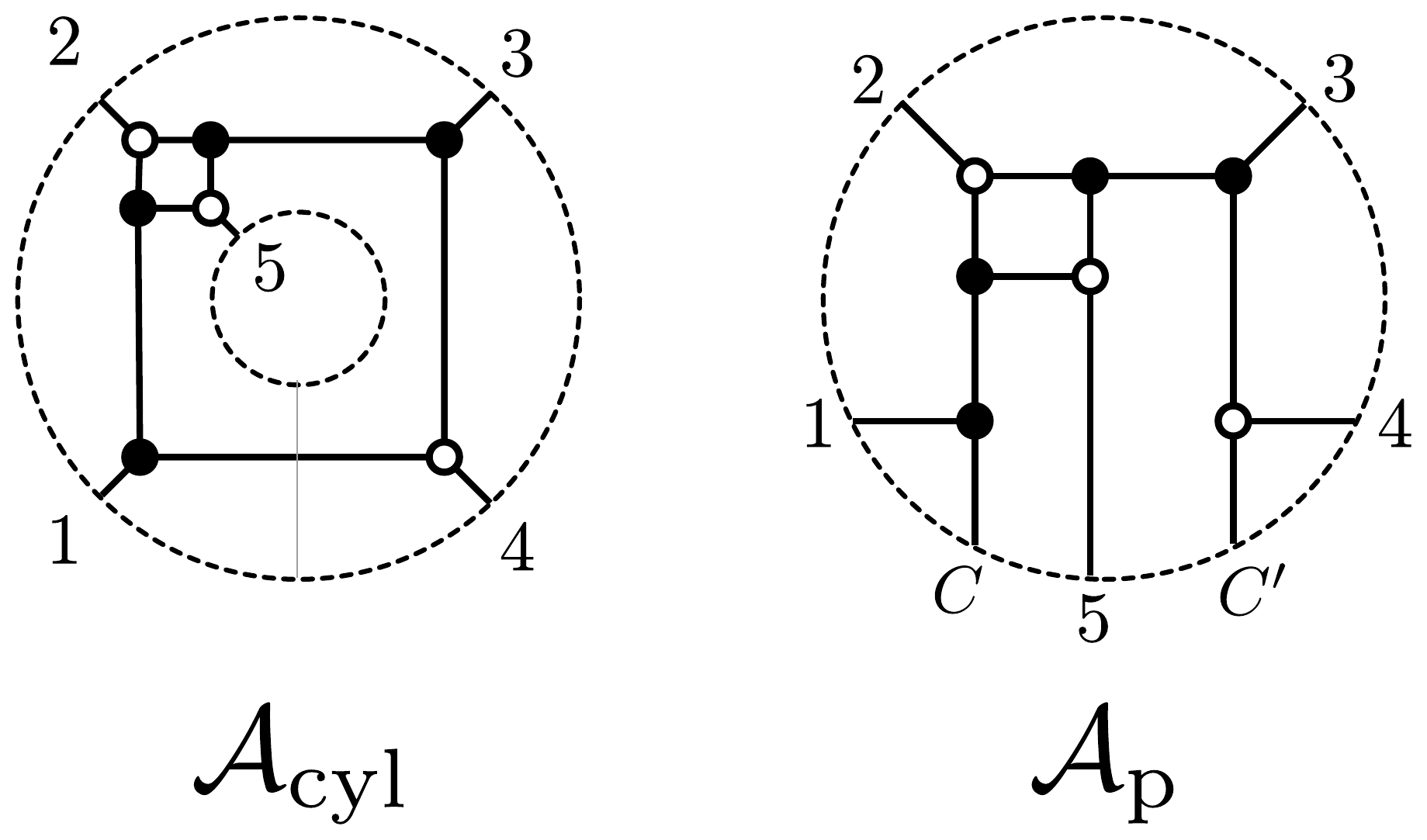}
    \caption{A five-point MHV diagram on a cylinder, and the planar diagram obtained after cutting along the indicated line.}
    \label{fig:example}
\end{figure}

\noindent
In this section we exemplify and validate the symmetries derived in Section~\ref{sec:2} and~\ref{sec:3} for a five-point MHV diagram with $\KKnp=2$ on a cylinder as depicted in Figure~\ref{fig:example}.  

While particle ``$5$'' belongs to one boundary $B'=(5)$, the remaining particles are on the other boundary $B=(1,2,3,4)$. As discussed in~\cite{Franco:2015rma}, the integrand of the Graßmannian integral corresponding to this diagram $\ampc$ can be written as
\begin{equation}\label{eq:intcyl}
    \Omega_{\mathrm{cyl}} = \frac{1}{(12)(23)(34)(41)} \, \frac{(13)}{(35)(51)}
    \eqndot
\end{equation} 
Here $(ij)$ denotes $2\times2$ minor with respect to the $i$th and $j$th column of the  $k\times n$ matrix ${\rm C}$, cf.~\eqref{eq:grassmannian}.

Let us first look at the action of the Yangian generators $\mono_B^{[i]\, ab}$, generated by the monodromy 
\begin{equation}
\mono_B(u) = \lax_1(u)\lax_2(u)\lax_3(u)\lax_4(u)
\eqncom 
\end{equation} 
on the particles at boundary $B$.
As discussed in Section~\ref{sec:2}, we can decompose the non-planar diagram $\ampc$ using the cutting procedure \eqref{eq:nplan}. 
When minimally cut, the planar diagram $\ampp$ has $\NNp=7$ external particles
and MHV degree $\KKp=3$ as shown in the right-hand side of Figure~\ref{fig:example}.
\noindent
Note that other possibilities to cut the diagram via a single edge are equivalent due to cyclic symmetry
on the boundary, while cutting two edges does not yield any identity.
As $n_B=4$, we find that the fourth level of the Yangian generators has to annihilate the non-planar on-shell function $\ampc$. The Yangian generators of this level read
\begin{equation}\label{eq:examplemono}
    \mono_B^{[4]\, ab}\!=(-1)^{ab+c+d+e}
    (\ww_4^a \dw_4^c)
    (\ww_3^c \dw_3^d)
    (\ww_2^d \dw_2^e)
    (\ww_1^e \dw_1^b)
 \eqndot
\end{equation}
In order to show that the $\ampc$ is annihilated by the operator above we proceed in analogy to~\cite{Drummond:2010qh}. After reordering \eqref{eq:examplemono} and acting on the delta function we obtain
\begin{multline}
  \!\!\!\!\!\mono_B^{[4]\, ab}\ampc=
  \\
  (-1)^{ab}\sum_{i=1}^4
    \int\!\!\frac{\dd^{2\times 5}{\rm C}}{\text{Vol}[GL(2)]} \, \Omega \, g(i)\ww_4^b \dw_i^a
    \delta^{8 | 8}({\rm C}\cdot\ww),
\end{multline} 
with 
$g(1)=\oo_{12}\oo_{23}\oo_{34}$,
$g(2)=-\oo_{23}\oo_{34}$,
$g(3)=\oo_{34}$,
$g(4)=-1$
and
$\oo_{ij}={\rm C}_{Ii}\frac{\partial}{\partial {\rm C}_{Ij}}$, where we sum over the index $I=1,2$.
Integrating by parts such that  the operators
$g(i)$  act only on $\Omega$, we find
\begin{equation}
    \mono_B^{[4]\, ab}\ampc=0
 \eqncom
\end{equation}
which agrees with \eqref{eq:resultyangian}.

We will now discuss the symmetries as derived in Section~\ref{sec:3}.
Since the diagram $\ampnp$ is embedded on a cylinder, the exact 
transfer matrix identity \eqref{eq:cylinderidentity} holds. In order to check this identity, we note that in our particular case there is only one particle on the boundary $B'$. 
Thus, due to the vanishing central charge constraint we trivially find
\begin{equation}\label{eq:exampletrans1}
 \trans_{B'}(u) \ampc =\str\lax_5(u) \ampc = 0
 \eqndot
\end{equation} 
The evaluation of the action of the transfer matrix on the particles at the boundary $B$ is more involved. However, again we can proceed in analogy to~\cite{Drummond:2010qh}. A straightforward calculation shows that 
\begin{multline}\label{eq:exampletrans}
 \!\!\!\!\trans_B(u) \ampc = \\ \sum_{i=0}^4 u^{4-i}\!\!\!\!\underbrace{\sum_{j_1>\ldots>j_i}\!\!
    (\ww_{j_1}^{a_1} \dw_{j_1}^{a_2})
    \cdots
    (\ww_{j_i}^{a_i} \dw_{j_i}^{a_1})}_{\trans^{[i]}_B} \ampc = 0
 \eqndot
\end{multline} 
Note that \eqref{eq:exampletrans} yields three independent identities when expanded in the spectral parameter $u$ for the action of $\trans^{[i]}_B$ with $i=2,3,4$, cf.~\eqref{eq:resulttransfer}. Here we did not include the case $i=0$ which identically vanishes as well as $i=1$ which trivially holds when acting on a function with vanishing central charge. The only identity that can be obtained from \eqref{eq:examplemono} by taking the supertrace is the case $i=4$.

The example also shows that one has to consider all possible embeddings in order
to find all symmetries, as briefly discussed at the end of Section~\ref{sec:2}. Note that in the diagram of $\ampc$ in Figure~\ref{fig:example}, another 
embedding on the cylinder is obtained by simply exchanging particles ``$2$'' and ``$5$''. The integrand \eqref{eq:intcyl} is invariant under this replacement.
Therefore the invariance relations \eqref{eq:examplemono} and \eqref{eq:exampletrans} also hold with 
the labels ``$2$'' and ``$5$'' interchanged.

\section{Conclusions}\noindent
In this paper we studied the integrable structures in non-planar contributions
to amplitudes in $\mathcal{N}\! = \! 4$ super Yang-Mills theory. 
More precisely, we have shown that leading singularities of non-planar amplitudes
in the maximally supersymmetric gauge theory in four dimensions enjoy Yangian-type symmetries. 
This is a direct consequence of the Yangian invariance of their planar counterparts.
We derived the action of the monodromy matrix, and thus the Yangian, as well as the corresponding transfer matrix on a given boundary of a given non-planar on-shell diagram.
The symmetry generators realize the same Yangian as in the planar case,
but act on each boundary of the diagram individually.
The lowest levels of these Yangian generators are broken, depending on
the minimal number of internal edges that have to be cut in order
to render the diagram planar.
Additionally, we showed that a similar statement also holds true for the commuting operators that arise from the expansion of the corresponding transfer matrix in the spectral parameter. 
As an example, we explicitly constructed the symmetry generators for a five-point MHV on-shell function on a cylinder.

It would be important to investigate if our prescription yields all unbroken generators when cutting a minimal number of internal edges. Also other ways of decomposing the non-planar diagrams or comparing different embeddings may yield further unbroken levels and additional constraints.
From a mathematical viewpoint, it would be interesting to investigate the algebra that results from combining the Yangian-type symmetries of the different boundaries with each other, and with the superconformal transformations which act on all external particles.

We hope that the symmetries can be used to determine integrands of non-planar Graßmannian integrals.
Despite the fact that the Yangian generators
are fairly complicated differential operators, their action on Graßmannian integrals can easily
be computed using the techniques outlined in Section~\ref{sec:example}. 
We have seen that in the example studied here the transfer matrix of length $n_B=4$ vanishes when acting on the on-shell function on the cylinder. Interestingly, the same is true when acting on the planar five-point MHV diagram. This suggests that both belong to the same $\mathfrak{gl}(4|4)$ multiplet. In principle, such a relation to planar diagrams may exist for all diagrams on the cylinder with a single leg on one boundary. It would be interesting to make this relation more precise and understand if it can be generalized to more involved configurations and topologies.

The most pressing question to ask is whether the symmetries we found manifest themselves in the full non-planar loop integrands and explain the recent observations which hint at hidden symmetries similar to the planar ones~\cite{Bern:2014kca,Arkani-Hamed:2014via,Bern:2015ple}. 
We plan to elaborate on these open problems elsewhere.

Finally, the key property of on-shell diagrams which allowed us to deduce their symmetries
from those of the planar ones
is the fact that for on-shell diagrams there is a well-defined
cutting and gluing procedure. As all internal states are on-shell, one can interpret them as external states of a cut diagram. 
Our formulation of the symmetries of on-shell diagrams indicates that it is possible
to systematically identify the implications of planar integrability for
the subleading terms in the $1/N$ expansion, for quantities where such a procedure 
is available. 
We hope that such a strategy can shed new light on the study of the integrable structure of non-planar observables, see e.g.~\cite{Kristjansen:2010kg}.
It would be interesting to connect these ideas to recent approaches 
to study amplitudes and correlation functions~\cite{Basso:2014hfa,Basso:2015zoa,Arkani-Hamed:2013jha,Bajnok:2015hla}.

\section*{Acknowledgements}\noindent
We thank Matthias Staudacher for inspiring discussions. Furthermore, we like to thank Patrick Dorey, Jan Fokken, Paul Heslop, Carlo Meneghelli, Arthur Lipstein, Brenda Penante and Jaroslav Trnka for fruitful discussions and comments on the manuscript.
RF likes to thank the \emph{Mathematical Physics of Space, Time and Matter} group at Humboldt-University Berlin for hospitality during multiple visits.
DM is supported by GK 1504 \emph{Masse, Spektrum, Symmetrie}.
The research leading to these results has received funding from the People Programme
(Marie Curie Actions) of the European Union’s Seventh Framework Programme FP7/2007-
2013/ under REA Grant Agreement No 317089 (GATIS).

\bibliographystyle{utphys2}
\renewcommand\refname{\normalsize References}
{\small
\bibliography{refs}
}

\end{multicols}
\end{document}